\documentclass[lettersize,journal]{IEEEtran}

\usepackage{algorithm}
\usepackage{algorithmic}

\usepackage{multirow}
\usepackage{booktabs} 
\usepackage{amsmath}
\usepackage{graphicx}
\usepackage{caption}

\usepackage{subcaption}
\usepackage{rotating}
\usepackage{adjustbox}
\usepackage{array}
\usepackage{float}
\usepackage{tabularx}
\usepackage{xcolor}
\usepackage{stfloats}
\usepackage{wrapfig}
\usepackage{listings}
\usepackage{color, colortbl}
\usepackage{balance} 
\usepackage{lscape}
\usepackage{xspace}
\usepackage{framed}
\usepackage{syntax}
\usepackage{parcolumns}
\usepackage{pifont}
\usepackage{soul}
\usepackage[tikz]{bclogo}
\usepackage{enumitem}
\usepackage{soul}
\usepackage{makecell}
\usepackage{hyperref}
\usepackage{tcolorbox}
\usepackage{enumitem}
\usepackage{ulem}
\usepackage{cite}

\newcommand{\figmargin}{\vspace{-15pt}}
\newcommand{\tabmargin}{\vspace{-10pt}}
\newcommand{\secmargin}{\vspace{-2pt}}

\newcommand{\boxmargin}{1mm}
\newtcolorbox{myboxc}{
    colback=gray!15!white,
    arc = 0pt, outer arc = 0pt,
    boxsep=0pt, left = 3pt, right = 0pt, top = 0pt, bottom = 0pt, 
    leftrule=3pt, bottomrule=0pt,toprule=0pt, rightrule=0pt,
    left = \boxmargin, right = \boxmargin, top = \boxmargin, bottom = \boxmargin
}

\newcommand{\myauthornote}[3]{}
\newcommand{\toolname}{RepoTransBench\xspace}

\newcommand{\newrevised}[1]{{\color{black} #1}}
\usepackage{authblk}

\begin{document}
\title{\toolname: A Real-World Multilingual Benchmark for Repository-Level Code Translation}

\author[1]{Yanli Wang}
\author[1,†]{Yanlin Wang \thanks{† Yanlin Wang is the corresponding author.}}
\author[1]{Suiquan Wang}
\author[1]{Daya Guo}
\author[1]{Jiachi Chen}
\author[2]{John Grundy}
\author[3]{Xilin Liu}
\author[3]{Yuchi Ma}
\author[1]{Mingzhi Mao}
\author[4]{Hongyu Zhang}
\author[1]{Zibin Zheng}

\affil[1]{Sun Yat-sen University, China}
\affil[2]{Monash University, Australia}
\affil[3]{Huawei Cloud Computing Technologies Co., Ltd, China}
\affil[4]{Chongqing University, China}
\affil[*]{Contact Emails: \texttt{wangyli58@mail2.sysu.edu.cn}, \texttt{wangylin36@mail.sysu.edu.cn}}

\maketitle

\begin{abstract}
Repository-level code translation refers to translating an entire code repository from one programming language to another while preserving the functionality of the source repository. Many benchmarks have been proposed to evaluate the performance of such code translators. However, previous benchmarks mostly provide fine-grained samples, focusing at either code snippet, function, or file-level code translation. Such benchmarks do not accurately reflect real-world demands, where entire repositories often need to be translated, involving longer code length and more complex functionalities.
To address this gap, we propose a new benchmark, named \textbf{\toolname}, which is a real-world multilingual repository-level code translation benchmark featuring 1,897 real-world repository samples across 13 language pairs with automatically executable test suites.
Besides, we introduce \textbf{RepoTransAgent}, a general agent framework to perform repository-level code translation.
We evaluate both our benchmark's challenges and agent's effectiveness using several methods and backbone LLMs, revealing that repository-level translation remains challenging, where the best-performing method achieves only a 32.8\% success rate.
Furthermore, our analysis reveals that translation difficulty varies significantly by language pair direction, with dynamic-to-static language translation being much more challenging than the reverse direction (achieving below 10\% vs. static-to-dynamic at 45-63\%).
Finally, we conduct a detailed error analysis and highlight current LLMs’ deficiencies in repository-level code translation, which could provide a reference for further improvements. 
We provide the code and data at \url{https://github.com/DeepSoftwareAnalytics/RepoTransBench}.
\end{abstract}



\section{Introduction}
Code translation refers to translating code from one programming language to another while preserving the functionality of the source code~\cite{pan2024lost,sun2024survey,eniser2024towards,dou2024s}. 
Recently, large language models (LLMs) have demonstrated strong performance across various tasks.
However, the training of these models increasingly faces limitations regarding the availability of open-source code data. High-quality synthetic data has become essential for further improving LLMs' performance~\cite{austin2021program,nijkamp2022codegen}. By leveraging code translation, vast amounts of code written in various programming languages can be transformed into high-quality training data, offering a viable solution to this scarcity~\cite{xie2023data,ren2020codebleu,hendrycks2021measuringcodingchallengecompetence}.
Beyond data synthesis, code translation has broad applications in other areas. It facilitates the refactoring of code written in outdated languages~\cite{aggarwal2015using}, transitions from simpler but slower languages to more complex and faster ones~\cite{szafraniec2022code}, and supports programming language migration in software development~\cite{nguyen2013lexical,nguyen2014migrating,mossienko2003automated,hassan2005lightweight,bartolomei2010swing,zhong2010mining,nguyen2014statistical,gu2017deepam}. These capabilities highlight the significance of code translation in addressing diverse challenges across the software engineering domain.
Automatic code translation can significantly reduce manual effort and has thus garnered widespread attention in recent years~\cite{bahdanau2014neural, chen2018tree, artetxe2018unsupervised, devlin2018bert, feng2020codebert, guo2020graphcodebert, ahmad2021unified, guo2022unixcoder, zheng2023codegeex}. With the popularity of large language models (LLMs), researchers are trying to translate code with LLMs, yielding promising results~\cite {yang2024UniTrans,lano2024using,nitin2024spectra}. 

To evaluate the performance of code translation tools, various benchmarks have been introduced~\cite{Zhu2022CoST,zhu2022xlcost,lachaux2020transcoder,lu2021codexglue,humanevalx,puri2021codenet,ahmad2021avatar,yan2023codescope,yan2023codetransocean}. 
Based on translation granularity, current fine-grained code translation benchmarks can be classified into three levels~\cite{pan2024lost}: 
snippet-level, function-level, and file-level. 
Specifically, \textbf{snippet-level} code translation benchmarks, such as CoST~\cite{Zhu2022CoST}, XLCost~\cite{zhu2022xlcost}, typically focus on evaluating the translation of program segments located between two consecutive code comments.
\textbf{Function-level} code translation benchmarks, such as TransCoder-test~\cite{lachaux2020transcoder}, CodeXGLUE~\cite{lu2021codexglue} and HumanEval-X~\cite{humanevalx}, focus on evaluating the translation of a function. \textbf{File-level} code translation benchmarks which include CodeNet~\cite{puri2021codenet}, Avatar~\cite{ahmad2021avatar}, CodeScope~\cite{yan2023codescope} and CodeTransOcean~\cite{yan2023codetransocean} refer to evaluating the translation of a complete program file.
However, these fine-grained code translation benchmarks may not meet the demands of real development scenarios, which require the translation of entire repositories.
Recently, Pan et al.~\cite{pan2024lost} manually study two open-source repositories (Apache Commons CLI~\cite{apachecommonscli} and Python Click~\cite{click}), and find that current LLMs struggle to perform code translation for entire repositories. 
Although this work has conducted preliminary research on repository-level code translation, we identify the following problems:

\begin{itemize}[left=2pt]
    \item \textbf{P1: Lack of Large-Scale Multilingual Benchmark.}
    Existing repository-level translation studies are severely limited in scope, with most work examining only one or two translation pairs and a handful of repositories. This narrow coverage fails to capture the diverse challenges posed by different programming paradigms, syntax variations, and ecosystem differences that occur across the broader landscape of programming languages used in real-world development.
    \item \textbf{P2: Labor-Intensive Execution-Based Test Suites Construction.} Repository-level code translation benchmark construction is labor-intensive, requiring extensive manual effort to ensure repository executability, generate comprehensive test suites, validate functional correctness, and manage complex dependencies.
    \item \textbf{P3: Lack of General Translation Framework for Different Translation Pairs.} Current translation methods often rely on language-specific heuristics tailored to particular translation pairs, making them difficult to generalize across diverse programming language combinations.
    \item \textbf{P4: Potentially Ignoring the Meta Information of Repositories.} Many programming language repositories contain configuration files and resource files, such as \texttt{CMakeLists.txt} for C++ and \texttt{pom.xml} for Java. A real-world code repository migration often requires proper handling of resources and correct configuration management.
\end{itemize}

In this paper, we introduce a \textbf{\textit{multilingual repository-level code translation benchmark}}, named \textbf{\toolname}, and a general \textbf{agent-based translation framework}, named \textbf{RepoTransAgent}, to address these limitations.

\toolname~ encompasses 1,897 repository samples across 13 translation pairs covering 7 programming languages, providing automatic execution-based test suites to evaluate both \textbf{compilability and functional correctness} of translated repositories. To construct \toolname, we develop a multi-agent framework that automatically generates comprehensive test suites and handles the complex requirements of repository-level translation validation. Our benchmark demonstrates higher complexity than previous work, with repositories containing an average of 23,966 tokens, 2,394 lines of code, 177 functions, 35 classes, and 163 import statements.

RepoTransAgent addresses general translation challenges through an intelligent agent framework based on the ReAct (Reasoning + Acting) paradigm, specifically designed to solve repository-level code translation problems. The agent iteratively combines reasoning about repository structure and translation requirements with concrete actions to handle the complexity of entire software projects. 
RepoTransAgent can analyze repository structures, understand cross-file dependencies, manage configuration files, and iteratively refine translations based on execution feedback. The agent operates through five core capabilities: reading files, creating files, executing commands, searching content, and marking completion, enabling repository-level translation through a reasoning-action loop that adapts to diverse programming language ecosystems.

We evaluate both the benchmark's challenges and our agent's effectiveness using several methods and backbone LLMs. Our experimental results reveal that repository-level code translation remains challenging for current methods, with the best-performing method achieving only 32.8\% success rate. However, our RepoTransAgent framework consistently outperforms baseline approaches, demonstrating improvements of up to 21.5\% over the error feedback method.

The key contributions of this research are:

\begin{itemize}[left=10pt]
    \item We introduce a large-scale repository-level code translation benchmark named \textbf{\toolname~} covering 13 translation pairs with 1,897 samples and automatic execution-based test suites. \toolname~ demonstrates substantially higher context and dependency complexity than previous benchmarks.
    \item We develop a multi-agent framework for automated benchmark construction to handle the complex requirements of repository-level translation validation and obtain the corresponding execution-based test suites.
    \item We propose \textbf{RepoTransAgent}, a general agent framework for multilingual repository-level code translation based on reasoning and action paradigms, achieving up to 32.8\% success rate on \toolname.
    \item We conduct an extensive evaluation across multiple dimensions, revealing that translation difficulty varies by different translation pairs and project complexity.
\end{itemize}

\begin{figure*}
    \centering
    \includegraphics[width=\linewidth]{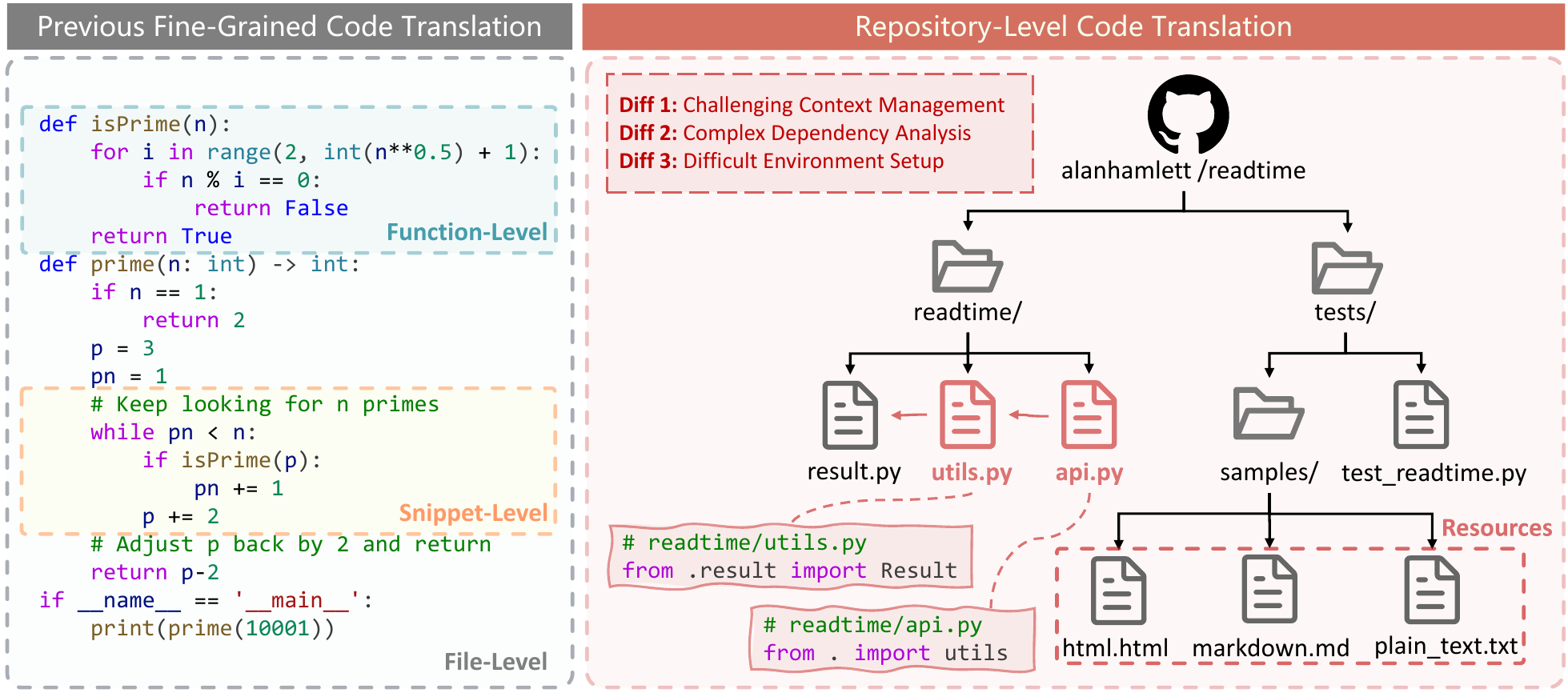}
    \caption{Comparison of different code translation granularity.}
    \label{fig:TranslationLevel}
\end{figure*}

\section{Background}

Code translation involves converting source code written in one programming language into another language while preserving the original program's functionality and logic~\cite{pan2024lost,sun2024survey,eniser2024towards,dou2024s}. This process is essential in software engineering for several reasons, such as migrating legacy systems to modern languages~\cite{nguyen2013lexical,nguyen2014migrating}, improving code performance by translating to more efficient languages~\cite{aggarwal2015using} and enabling cross-platform compatibility~\cite{szafraniec2022code}. As programming languages continue to evolve, the demand for accurate and efficient code translation techniques has grown, making it a critical area of research and development~\cite{bahdanau2014neural, chen2018tree, artetxe2018unsupervised, devlin2018bert, feng2020codebert, guo2020graphcodebert, ahmad2021unified, guo2022unixcoder, zheng2023codegeex}. The field has experienced significant technological evolution, with each advancement enabling translation capabilities at increasingly larger code granularities. \newrevised{The development of code translation can be broadly divided into three main approaches based on their underlying technologies: Rule-Based Translation, Neural Network-Based Translation, and Large Language Model-Based Translation.}

\newrevised{\textbf{Rule-Based Translation.}}
\newrevised{The earliest code translation approaches relied on manually defined rules and grammar specifications.} Existing non-learning-based code translation techniques can be categorized into several main groups. Parser-based tools like ANTLR~\cite{ANTLR} rely on manually defined grammar rules to translate source code between languages. Transpilers such as Babel~\cite{Babel}, Emscripten~\cite{Emscripten}, JSweet~\cite{JSweet}, and GWT~\cite{GWT} convert source code from one language to another, often used to ensure compatibility across platforms or systems. Domain-specific translators like CxGo~\cite{CxGo}, C2Rust~\cite{c2rust}, and JavaToCSharp~\cite{javaTocsharp} focus on specific translation pairs, offering targeted translation solutions. Intermediate language compilers like Haxe~\cite{Haxe} compile code to a variety of target languages by using an intermediate format. Interface generators such as SWIG~\cite{SWIG} create cross-language bindings, allowing different languages to interact with each other without direct code translation. 
\newrevised{While these rule-based approaches demonstrate high precision for well-defined translation patterns, they face significant limitations in handling complex code structures and require substantial manual effort to maintain and extend.} Most rule-based methods primarily support \textbf{snippet-level} code translation~\cite{Zhu2022CoST,zhu2022xlcost}, which typically refers to evaluating the translation of program segments that are located between two consecutive code comments, and each program may consist of one or more code snippets. Some advanced rule-based tools can handle simple \textbf{function-level} translation where functions can be processed independently without complex external dependencies.

\newrevised{\textbf{Neural Network-Based Translation.}}
\newrevised{The introduction of neural networks marks a paradigm shift in code translation, enabling more sophisticated translation capabilities through learned representations.} Early learning-based approaches often train a neural network to achieve the ability of code translation~\cite{feng2020codebert,guo2020graphcodebert,wang2021codet5,guo2022unixcoder}.
Aggarwal et al.~\cite{aggarwal2015using} convert Python 2 code to Python 3 code using trained Moses~\cite{koehn2007moses}, which is an open-source toolkit for statistical machine translation.
Chen et al.~\cite{chen2018tree} design a tree-to-tree neural network to translate a source tree into a target one. DeepAM~\cite{gu2017deepam} discusses the limitations of bilingual projects, as well as the automatic mining of API mappings to reduce manual effort in code migration. 
Zheng et al.~\cite{zheng2017maximum} propose an approach for zero-resource NMT using maximum expected likelihood estimation.
TransCoder~\cite{lachaux2020transcoder} is a transformer with 6 layers to perform code translation at function level.
Besides, some models pre-trained on multilingual corpora like Codex~\cite{chen2021evaluating}, CodeT5~\cite{wang2021codet5}, and CodeGen~\cite{nijkamp2022codegen} demonstrate remarkable code translation capability.
These neural approaches successfully enable practical \textbf{function-level} translation~\cite{lachaux2020transcoder,Zhu2022CoST,zhu2022xlcost,humanevalx,lu2021codexglue}, which refers to translating a function into another programming language, with the data sources often being manually crafted datasets~\cite{humaneval} or coding practice websites~\cite{geeksforgeeks}. More advanced neural systems achieve \textbf{file-level} code translation~\cite{puri2021codenet,ahmad2021avatar,khan2023xcodeeval,yan2023codescope,yan2023codetransocean,yin2024rectifier}, which often refers to translating a complete program file into the target language. The data sources are usually from code contest platforms~\cite{Codeforces,atcoder,aizu,GoogleCodeJam} or task solutions websites~\cite{dotnetsamples,d2lai,rosettacode}. G-TransEval~\cite{jiao2023GTransEval} also provides a more fine-grained taxonomy, including token-level, syntax-level, library-level, and algorithm-level, which is part of a function. However, neural network-based approaches still struggle with cross-file dependencies and large-scale context management, limiting their effectiveness beyond file-level translation.

\newrevised{\textbf{Large Language Model-Based Translation.}} 
In recent years, large language models 
such as 
StarCoder~\cite{li2023starcoder,lozhkov2024starcoder}, SantaCoder~\cite{allal2023santacoder} and more latest models such as the Llama series~\cite{llama,llama2,llama3}, the ChatGPT series~\cite{openai2023gpt4}, the DeepSeek series~\cite{deepseekv2}, and the Claude series~\cite{Claude} have shown remarkable performance on traditional code translation tasks. These models are trained on large code corpus and have strong comprehension and instruction following abilities, 
which can perform accurate and efficient code translations on previous fine-grained code benchmarks. 
Recent research has explored various techniques to enhance LLM-based translation capabilities. Rectifier~\cite{yin2024rectifier} is a fine-tuned micro model that acts as a general corrector to correct the translation errors of unknown LLMs. Vert~\cite{yang2024vert} leverages LLM's strong few-shot learning ability to 
produce readable Rust translations with formal guarantees of correctness.
Bhattarai et al.~\cite{bhattarai2024enhancing} enhance code translation in LLMs with few-shot learning via retrieval-augmented generation. 
TransAgent~\cite{yuan2024transagent} is an LLM-based multi-agent system for code translation.
SpecTra~\cite{nitin2024spectra} considers the different kinds of specifications that can be extracted from a program to enhance the code translation ability of LLMs. Momoko et al.~\cite{shiraishi2024context} propose an LLM-based translation scheme that improves the success rate of translating large-scale C code into compilable Rust code. SolMover~\cite{karanjai2024teaching} can convert smart contracts written in Solidity~\cite{solidity} to Move~\cite{move} with LLMs. 
CCTrans~\cite{yang2024cctrans} can transpile concurrent Java files to JavaScript using multiple workers while maintaining identical behavior. 
GlueTest~\cite{abidgluetest} systematically and semiautomatically validates translations for non-trivial libraries. 
UniTrans~\cite{yang2024UniTrans} is a unified code translation framework applicable to various LLMs to unleash their power in code translation. SDA-Trans~\cite{liu2023syntax} is a syntax and domain-aware model for program translation, which leverages the syntax structure and domain knowledge to enhance the cross-lingual transfer ability.

Figure~\ref{fig:TranslationLevel} shows a comparison of code granularities used in different code translation approaches across these technological eras. Most previous works focus on the translations with a granularity not exceeding a single code file (left-hand side). Unlike the previous fine-grained granularity code translation, \textbf{repository-level} code translation involves migrating an entire repository from one language to another. Recently, repository-level code translation has gradually gained the attention of researchers~\cite{pan2024lost}. As shown in the right-hand side of Figure~\ref{fig:TranslationLevel}, a typical code repository contains functional code files and test code files and may also include resource files and configuration files. Functional code files refer to those code files that implement specific functionalities of the code repository, such as the files located in the \texttt{readtime/} directory. Test code files refer to the files used to verify the correctness of the functional code where \texttt{test_readtime.py} is an example. Resource files like those in \texttt{samples/} folder are used to test the functional correctness of the functional code. The functional code needs to implement how to perform I/O operations with these resources. In addition, for certain language-specific frameworks, it is necessary to complete the configuration file correctly, such as the ``pom.xml'' file in Java's Maven~\cite{maven} repositories.

Compared with previous fine-grained code translation granularity, repository-level code translation presents three fundamentally different challenges: challenging context management, complex dependency analysis, and difficult environment setup. Real-world code repositories typically include numerous functions, classes, and import statements to realize complex functionalities, requiring translators to understand the entire repository context rather than isolated code fragments with intricate interdependencies between components across multiple files and modules. The complexity extends beyond code volume to sophisticated dependency management requirements, where files like \texttt{api.py} and \texttt{utils.py} have complex import relationships (e.g., \texttt{from .result import Result}, \texttt{from . import utils}) that must be correctly analyzed and maintained during translation. Additionally, successful repository translation necessitates appropriate configuration of ecosystem-specific files and comprehensive resource migration beyond source code, where resource files that perform I/O operations with code components must be carefully handled and potentially transformed to maintain functional equivalence.
Furthermore, unlike artificially crafted datasets used in fine-grained translation~\cite{ahmad2021avatar,puri2021codenet}, real repositories typically contain existing test suites that can serve as valuable validation mechanisms. Effective repository-level translation must leverage these test cases not only for validation but also as specifications for maintaining functional correctness throughout the translation process.

Recent studies have shown interest in repository-level code translation. Pan et al.~\cite{pan2024lost} attempt to perform mutual conversion between Python and Java projects~\cite{apachecommonscli,click}, but find that the advanced LLMs are largely ineffective, with success rates of 8.1\% for GPT-4 and 0\% for the rest of the models. This stark performance gap reveals that repository-level translation requires fundamentally different approaches from fine-grained translation, as current LLM-based methods struggle with the challenges of large-scale context management, complex dependency analysis, and comprehensive environment configuration that are inherent to repository-level code translation.

\section{RepoTransBench}

\begin{figure*}
    \centering
    \includegraphics[width=0.95\linewidth]{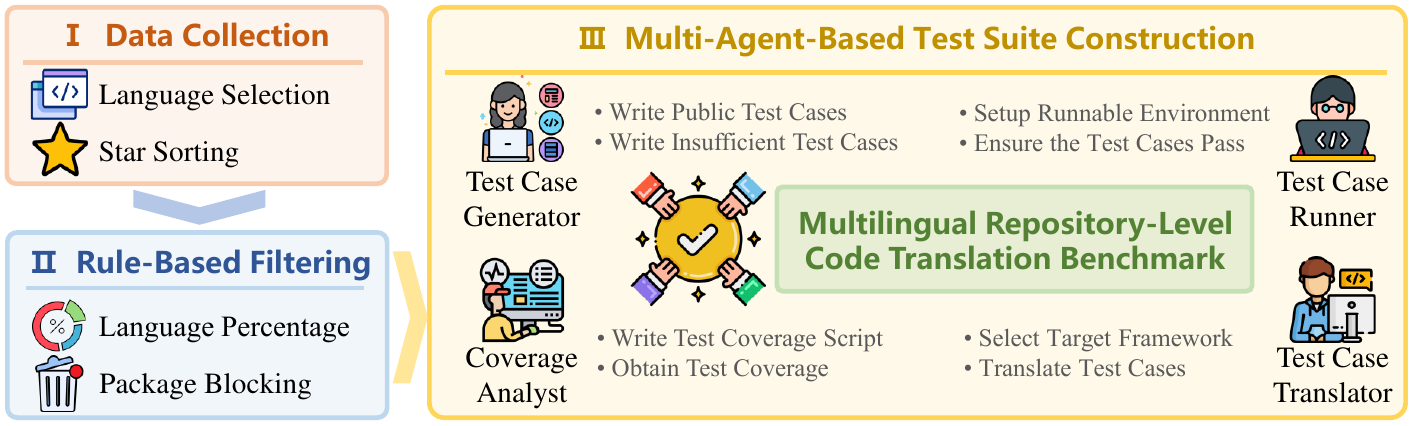}
    \caption{RepoTransBench construction pipeline.}
    \label{fig:BenchmarkConstruction}
\end{figure*}

As illustrated in Figure~\ref{fig:BenchmarkConstruction}, the data collection pipeline of RepoTransBench consists of three steps: data collection, rule-based filtering, and multi-agent-based test suite construction.

\subsection{Data Collection}

We conduct a questionnaire among professional developers to identify practical demands for repository-level code translation. We select the top 7 languages from the TIOBE programming language rankings~\cite{TIOBE-Index} selecting Rust and Matlab as candidate languages for our translation pair matrix. Additionally, we provide custom options for respondents to specify other translation pairs they require. 
The detailed questionnaire and statistical results are available in the artifact.

We receive responses from 21 professional developers, comprising 86 requests for specific translation pairs. The survey results reveal diverse translation needs. The highest demand is observed for Python$\leftrightarrow$C++ bidirectional translation, with developers frequently needing to migrate between Python's rapid prototyping capabilities and C++'s performance-critical applications in systems programming and high-performance computing. JavaScript$\rightarrow$Python translation also shows strong demand as organizations seek to consolidate their tech stacks by moving web-based logic into Python's rich ecosystem for data processing and machine learning workflows.

The sustained interest in Python$\rightarrow$Java translation reflects the common enterprise pattern where Python prototypes must be productionized in Java environments for scalability and integration with existing enterprise systems. Meanwhile, C$\rightarrow$Rust translation demand indicates the growing adoption of Rust for systems programming, where developers aim to modernize legacy C codebases while gaining memory safety guarantees. The interest in Python$\rightarrow$Go translation similarly reflects organizational shifts toward Go's concurrency model and deployment simplicity for backend services.

Repository-level translation becomes particularly valuable in these scenarios because modern software projects involve complex interdependencies, build systems, and architectural patterns that cannot be addressed through isolated function or class translations. Developers require tools that can maintain semantic correctness across entire codebases while preserving project structure and dependency relationships.

Based on these survey findings, we expand our benchmark to include 13 translation pairs that reflect real-world developer needs. To ensure repository quality, we rank repositories by star count and retain only those with more than 50 stars.

\subsection{Rule-Based Filtering}

\newrevised{We provide a specification of our filtering rules in Algorithm~\ref{alg:filtering}. The rule-based filtering implements three key criteria:
\textbf{Rule 1: Language Dominance.} We retain only repositories where the target language constitutes the largest proportion of the codebase. This ensures that the repository is primarily written in the language we aim to translate.
\textbf{Rule 2: Popularity Threshold.} We filter repositories based on a minimum star threshold $\theta_{star}$ to ensure code quality and practical relevance.
\textbf{Rule 3: Package Exclusion.} We exclude repositories that depend on packages difficult to translate into target languages. The detailed package lists are available at~\cite{repotransbenchURL}.}
To illustrate, for Python repositories, we require that (1) Python code comprises the largest share of the codebase, (2) the repository exceeds the popularity threshold, and (3) it does not have substantial dependencies on excluded packages (e.g., PyTorch, TensorFlow, NLTK).

\begin{algorithm}[h]
\caption{\newrevised{Rule-Based Repository Filtering}}
\label{alg:filtering}
\begin{algorithmic}[1]
\setlength{\lineskip}{0pt}
\setlength{\baselineskip}{10pt}
\REQUIRE Collection of repositories $\mathcal{R} = \{R_1, R_2, \ldots, R_n\}$, Language $L_{target}$, Star threshold $\theta_{star}$
\ENSURE Filtered repository collection $\mathcal{R}_{filtered} \subseteq \mathcal{R}$
\STATE \textbf{function} \textsc{PassesAllRules}($R, L_{target}, \theta_{star}$)
\STATE \quad // Rule 1: Language Dominance
\IF{$P(L_{target}, R) \neq \max\{P(L_i, R) \mid L_i \in \text{Languages}(R)\}$}
    \STATE \quad \textbf{return} false
\ENDIF
\STATE \quad // Rule 2: Popularity Threshold
\IF{$\text{Stars}(R) < \theta_{star}$}
    \STATE \quad \textbf{return} false
\ENDIF
\STATE \quad // Rule 3: Package Blocking
\IF{$R \cap \text{BlockedPackages}(L_{target}) \neq \emptyset$}
    \IF{\textbf{not} $\text{IsBasicUsage}(R)$}
        \STATE \quad \quad \textbf{return} false
    \ENDIF
\ENDIF
\STATE \quad \textbf{return} true
\STATE \textbf{end function}
\STATE
\STATE $\mathcal{R}_{filtered} \leftarrow \emptyset$
\FOR{each $R \in \mathcal{R}$}
    \IF{$\textsc{PassesAllRules}(R, L_{target}, \theta_{star})$}
        \STATE $\mathcal{R}_{filtered} \leftarrow \mathcal{R}_{filtered} \cup \{R\}$
    \ENDIF
\ENDFOR
\STATE \textbf{return} $\mathcal{R}_{filtered}$
\end{algorithmic}
\end{algorithm}

\subsection{Multi-Agent-Based ~\newrevised{Test Suite} Construction}

Due to the substantial human resources and effort required for benchmark construction, and to facilitate scalable expansion of dataset size and translation pair types, we develop a multi-agent framework for constructing execution-based repository-level code translation benchmarks. The framework comprises four specialized agents that work collaboratively to ensure high-quality benchmark construction: Test Case Generator, Test Case Runner, Coverage Analyst, and Test Case Translator.

\textbf{Test Case Generator} analyzes the source repository structure and functionality to generate comprehensive test cases. It performs two primary functions: (1) writing public test cases that cover the main functionality and API interfaces of the repository, ensuring that critical code paths are exercised, and (2) identifying insufficient test cases by analyzing code coverage gaps and generating additional test scenarios to improve overall test completeness.

\textbf{Test Case Runner} focuses on environment setup and test execution validation. This agent is responsible for: (1) setting up runnable environments by analyzing project dependencies, installing required packages, and configuring build systems (such as Maven for Java projects or pip for Python projects), and (2) ensuring all test cases pass successfully in the source language environment. 

\textbf{Coverage Analyst} provides a quantitative assessment of test quality through comprehensive coverage analysis. This agent: (1) writes language-specific test coverage scripts tailored to each programming language's testing frameworks and coverage tools, and (2) obtains detailed test coverage metrics including line coverage, branch coverage, and function coverage. The coverage analysis ensures that our benchmark maintains high-quality standards, with our current dataset achieving an average of 81.89\% line coverage and 72.61\% branch coverage across all translation pairs as shown in Table~\ref{table:languageStats}.

\textbf{Test Case Translator} handles the cross-language translation of test cases to ensure translated repositories can be properly validated. This agent: (1) selects appropriate target frameworks by analyzing the functionality requirements and identifying equivalent libraries and testing frameworks in the target language, and (2) translates test cases from the source language to the target language while preserving test semantics and assertions. The agent maintains a mapping of equivalent libraries and frameworks across different programming languages to ensure translated test cases accurately reflect the original test intentions.

These agents can perform various operations, including file reading/writing, directory traversal, package installation, search operations, and executing command-line instructions. Through multi-agent collaboration and iterative refinement, the framework operates in a coordinated pipeline where each agent's output serves as input for subsequent agents, ultimately ensuring that source code repositories are executable and successfully generating corresponding test cases in target languages.

\subsection{Statistics of RepoTransBench}

\begin{table*}[t]
  \centering
  \footnotesize
  \setlength\tabcolsep{1.5pt}
  \caption{Statistics of \toolname~ compared to other existing code translation datasets. Considering the translation of Python to Java, we report the number of samples and the average number of tokens, lines, functions, classes, and import statements per sample of each task. \#Tokens counts are based on OpenAI’s tiktoken tokenizer (\url{https://github.com/openai/tiktoken}). \#Funcs, \#Classes and \#Imports counts are based on tree-sitter (\url{https://tree-sitter.github.io}). Code comments are removed before computation.}
\resizebox{\linewidth}{!}{
    \begin{tabular}{l|c|c|c|c|c|c|c|c|c|c|c}
    \hline
    \multicolumn{1}{c|}{\textbf{Dataset}} & \multicolumn{1}{c|}{\textbf{Year}} & \multicolumn{1}{c|}{\textbf{Source}} & \multicolumn{1}{c|}{\textbf{Level}} & \multicolumn{1}{c|}{\textbf{Config File}} & \multicolumn{1}{c|}{\textbf{Evaluation}} & \multicolumn{1}{c|}{\textbf{\#Samples}} & \multicolumn{1}{c|}{\textbf{\#Tokens}} & \multicolumn{1}{c|}{\textbf{\#Lines}} & \multicolumn{1}{c|}{\textbf{\#Funcs}} & \multicolumn{1}{c|}{\textbf{\#Classes}} & \multicolumn{1}{c}{\textbf{\#Imports}} \\
    \hline
    \hline
    TransCoder-test~\cite{lachaux2020transcoder} & 2020 & GeeksforGeeks~\cite{geeksforgeeks} & Function & Not Req & Execution & 868 & 127 & 12 & 1 & 0 & 0 \\
    \hline
    CodeNet~\cite{puri2021codenet} & 2021 & AIZU~\cite{aizu}, AtCoder~\cite{atcoder} & File & Not Req & Execution & 200 & 99 & 12 & 1 & 0 & 0 \\
    \hline
    Avatar~\cite{ahmad2021avatar} & 2021 & AtCoder~\cite{atcoder}, etc\footnotemark[1] & File & Not Req & Execution & 250 & 175 & 18 & 1 & 0 & 1 \\
    \hline
    \multirow{2}{*}{CoST~\cite{Zhu2022CoST}} & \multirow{2}{*}{2022} & \multirow{2}{*}{GeeksforGeeks~\cite{geeksforgeeks}} & Snippet & Not Req & Similarity & 351 & 33 & 3 & 0 & 0 & 0 \\
    \cline{4-12}
     & &  & Function & Not Req & Similarity & 69 & 173 & 15 & 1 & 0 & 0 \\
    \hline
    \multirow{2}{*}{XLCoST~\cite{zhu2022xlcost}} & \multirow{2}{*}{2022} & \multirow{2}{*}{GeeksforGeeks~\cite{geeksforgeeks}} & Snippet & Not Req & Similarity & 6861 & 24 & 2 & 0 & 0 & 0 \\
    \cline{4-12}
     & &  & Function & Not Req & Similarity & 864 & 196 & 19 & 1 & 0 & 0 \\
    \hline
    HumanEval-X~\cite{humanevalx} & 2022 & HumanEval~\cite{humaneval} & Function & Not Req & Execution & 164 & 65 & 8 & 1 & 0 & 0 \\
    \hline
    G-TransEval~\cite{jiao2023GTransEval} & 2023 & HumanEval~\cite{humaneval}, etc\footnotemark[2] & Function\footnotemark[3] & Not Req & Similarity & 400 & 90 & 10 & 1 & 0 & 0 \\
    \hline
    xCodeEval~\cite{khan2023xcodeeval} & 2023 & Codeforces~\cite{Codeforces} & File & Not Req & Execution & 1942 & 209 & 22 & 1 & 0 & 1 \\
    \hline
    \hline
    CodeScope~\cite{yan2023codescope} & 2023 & Codeforces~\cite{Codeforces} & File & Not Req & Execution & 30 & 259 & 28 & 1 & 0 & 1 \\
    \hline
    CodeTransOcean~\cite{yan2023codetransocean} & 2023 & Rosetta Code~\cite{rosettacode}, d2l-ai~\cite{d2lai} & File & Not Req & Similarity & 1029 & 253 & 24 & 2 & 0 & 1 \\
    \hline
    UniTrans~\cite{yang2024UniTrans} & 2024 & GeeksforGeeks~\cite{geeksforgeeks} & Function & Not Req & Execution & 568 & 112 & 11 & 1 & 0 & 0\\
    \hline
    \hline
    \rowcolor{gray!20}
    \textbf{\toolname~} & \textbf{2024} & \textbf{GitHub} & \textbf{Repository} & \textbf{Require} & \textbf{Execution} & \textbf{1897} & \textbf{23966} & \textbf{2394} & \textbf{177} & \textbf{35} & \textbf{163} \\
    \hline
    \end{tabular}}%
 \label{table:datasets}
\tabmargin
\end{table*}%

\begin{table}[t]
  \centering
  \footnotesize
  \setlength\tabcolsep{2pt}
  \caption{Statistics of source languages showing cross-file/intra-file dependencies and test coverage. Src. Lang.: source language, Proj.: projects, Samp.: Samples, Cross.: cross-file dependencies, Intra.: intra-file dependencies, Cov.: coverage. Translation pairs: C$\rightarrow$\{Python, Rust\}, C++$\rightarrow$Python, C\#$\rightarrow$Java, Java$\rightarrow$\{C\#, Go, Python\}, JavaScript$\rightarrow$Python, Matlab$\rightarrow$Python, Python$\rightarrow$\{C++, Go, Java, Rust\}. Total: 13 pairs, 1897 translations.}
\resizebox{\linewidth}{!}{
    \begin{tabular}{l|c|c|c|c|c|c}
    \hline
    \multicolumn{1}{c|}{\textbf{Src. Lang.}} & \multicolumn{1}{c|}{\textbf{\#Proj.}} & \multicolumn{1}{c|}{\textbf{\#Samp}} & \multicolumn{1}{c|}{\textbf{\#Cross.}} & \multicolumn{1}{c|}{\textbf{\#Intra.}} & \multicolumn{1}{c|}{\textbf{Line Cov.}} & \multicolumn{1}{c}{\textbf{Branch Cov.}} \\
    \hline
    \hline
    C & 122 & 244 & 64.3 & 2494.4 & 91.71\% & 63.59\% \\
    \hline
    C++ & 181 & 181 & 94.9 & 2401.3 & 86.20\% & 58.98\% \\
    \hline
    C\# & 97 & 97 & 31.1 & 600.9 & 83.52\% & 80.21\% \\
    \hline
    Java & 146 & 438 & 156.9 & 1286.5 & 73.21\% & 66.86\% \\
    \hline
    JavaScript & 189 & 189 & 27.0 & 707.5 & 93.56\% & 88.12\% \\
    \hline
    Matlab & 64 & 64 & 2.1 & 1609.5 & 61.33\% & 55.99\% \\
    \hline
    Python & 171 & 684 & 128.1 & 1006.1 & 81.29\% & 79.26\% \\
    \hline
    \hline
    \rowcolor{gray!20}
    \textbf{Overall} & \textbf{970} & \textbf{1897} & \textbf{87.9} & \textbf{1443.7} & \textbf{81.89\%} & \textbf{72.61\%} \\
    \hline
    \end{tabular}}%
 \label{table:languageStats}
\tabmargin
\end{table}%

Tables~\ref{table:datasets} and~\ref{table:languageStats} present comprehensive statistics of RepoTransBench. Table~\ref{table:datasets} compares RepoTransBench with existing code translation benchmarks, while Table~\ref{table:languageStats} provides detailed statistics across different source languages in our benchmark. Our benchmark represents a paradigm shift from previous fine-grained approaches to repository-level translation, demonstrating unprecedented scale and complexity in code translation evaluation.

\textbf{Scale and Length Comparison.} As shown in Table~\ref{table:datasets}, unlike previous benchmarks that operate at function, snippet, or single-file levels, RepoTransBench operates at the repository-level, encompassing complete software projects with their inherent complexity. Our benchmark contains 1,897 translation samples across 13 translation pairs, significantly surpassing the scale of previous work. The average sample in our benchmark contains 23,966 tokens and 2,394 lines of code, representing a dramatic increase of over 95× in tokens and 108× in lines compared to the largest previous benchmark (xCodeEval with 209 tokens and 22 lines per sample).

\textbf{Structural Complexity.} As shown in Table~\ref{table:datasets}, the repository-level nature of our benchmark introduces structural elements absent in previous work. Each sample contains an average of 177 functions, 35 classes, and 163 import statements, demonstrating the complex interdependencies and architectural patterns inherent in real-world software projects. In contrast, previous benchmarks typically contain at most 2 functions per sample and rarely include classes or substantial import dependencies.

\newrevised{\textbf{Dependency Complexity.} As shown in Table~\ref{table:languageStats}, RepoTransBench exhibits substantial dependency complexity that distinguishes it from fine-grained benchmarks. RepoTransBench has an average of 87.9 cross-file dependencies and 1,443.7 intra-file dependencies. The diversity in dependency complexity across different programming languages reflects varied architectural patterns: cross-file dependencies range from 2.1 in Matlab to 156.9 in Java, while intra-file dependencies range from 600.9 in C\# to 2,494.4 in C. The complexity distinguishes repository-level translation from function-level or file-level translation, where such cross-module dependencies are absent or minimal.}

\textbf{Configuration Requirements.} As shown in Table~\ref{table:datasets}, another crucial distinction of RepoTransBench is the requirement for proper configuration files (such as ``pom.xml'' for Java Maven projects, ``package.json'' for JavaScript projects, or ``requirements.txt'' for Python projects). This requirement reflects real-world translation scenarios where successful code migration depends not only on translating source code but also on correctly configuring build systems, dependency management, and project structure in the target language ecosystem.

\textbf{Cross-Language Coverage.} Table~\ref{table:languageStats} provides detailed statistics across our 13 supported translation pairs, covering translations from C, C++, C\#, Java, JavaScript, Matlab, and Python to various target languages. Our dataset encompasses 970 unique projects, with Python being the most represented source language (684 samples from 171 projects) followed by Java (438 samples from 146 projects).

\textbf{Test Coverage Quality.} Our multi-agent framework ensures high-quality test coverage across all translation pairs. As shown in Table~\ref{table:languageStats}, the overall dataset maintains 81.89\% line coverage and 72.61\% branch coverage, with JavaScript achieving the highest coverage (93.56\% line coverage, 88.12\% branch coverage) and Matlab showing the most room for improvement (61.33\% line coverage, 55.99\% branch coverage). This comprehensive test coverage enables reliable execution-based evaluation of translation quality, moving beyond similarity metrics used in previous benchmarks to assess actual functional correctness.

\newrevised{\textbf{Translation Pair Classification.} We classify the 13 translation pairs from the type system perspective, which can be categorized into four groups: static-to-dynamic translations (C/C++/Java$\rightarrow$Python, Matlab$\rightarrow$Python), dynamic-to-static translations (Python$\rightarrow$C++/Java/Go/Rust), static-to-static translations (C$\rightarrow$Rust, C\#$\rightarrow$Java, Java$\rightarrow$C\#/Go), and dynamic-to-dynamic translation (JavaScript$\rightarrow$Python).
}

\section{RepoTransAgent}

\begin{figure*}
    \centering
    \includegraphics[width=0.9\linewidth]{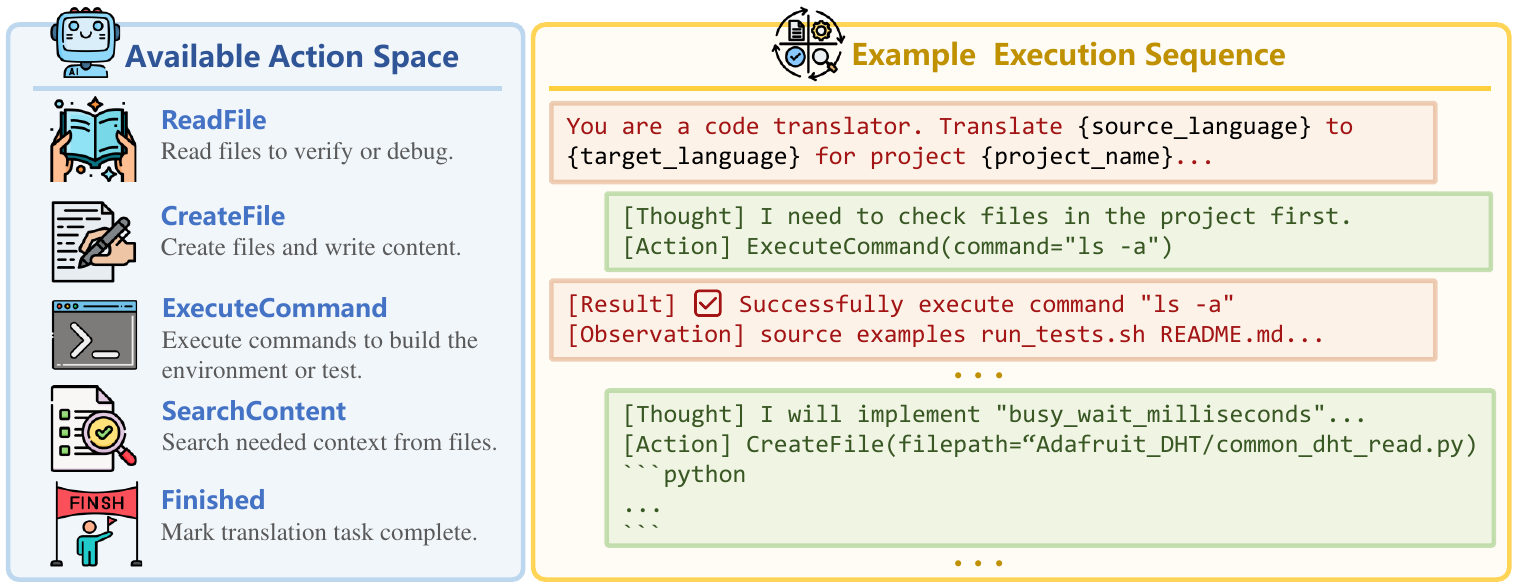}
    \caption{Overview of the RepoTransAgent.}
    \label{fig:RepoTransAgent}
\end{figure*}

We propose RepoTransAgent, a general agent framework designed for repository-level code translation. RepoTransAgent adopts a holistic understanding of entire software repositories, enabling it to handle complex interdependencies, maintain project structure, and ensure functional correctness during translation.

\subsection{Agent Architecture and Action Space}

As illustrated in Figure~\ref{fig:RepoTransAgent}, RepoTransAgent operates through an action space consisting of five core capabilities that enable repository analysis and translation:

\textbf{ReadFile} allows the agent to examine existing code files, configuration files, documentation, and dependency specifications. This action is crucial for understanding the repository structure, identifying key components, and analyzing code patterns. The agent uses this capability to verify translation correctness and debug issues during the translation process.

\textbf{CreateFile} enables the agent to generate new files in the target language, including code files, configuration files (such as \texttt{pom.xml} for Java projects or \texttt{package.json} for JavaScript projects), and build scripts. This action is essential for establishing the translated repository structure and ensuring proper project organization in the target language ecosystem.

\textbf{ExecuteCommand} provides the agent with the ability to execute system commands for environment setup, dependency installation, compilation, and testing. This capability enables the agent to validate translations by running build processes and executing test cases, ensuring that the translated repository maintains functional correctness.

\textbf{SearchContent} allows the agent to efficiently locate specific code patterns, function definitions, class declarations, and import statements across the repository. This search capability is particularly important for understanding cross-file dependencies and ensuring consistent translation of related components throughout the project.

\textbf{Finished} marks the completion of the translation task, indicating that the agent has successfully translated the repository and validated its correctness through execution-based testing.

\subsection{Translation Process and Execution Workflow}

RepoTransAgent follows a translation workflow based on the ReAct paradigm that begins with comprehensive repository analysis and proceeds through iterative translation and validation cycles. At the start of each translation task, the agent is initialized as a code translator and provided with basic information about the current working directory and translation requirements, including the source language, target language, project name, and other relevant details.

The translation process initiates with the agent's reasoning phase, where it formulates a plan in the [Thought] section to analyze the repository structure systematically. As demonstrated in the example execution sequence, the agent first reasons that it needs to examine the files in the project directory, then executes the corresponding action using \texttt{ExecuteCommand(command="ls -a")} to explore the repository contents. The system provides feedback to the agent through two components: [Result] indicating the execution status (successfully executed in this case), and [Observation] containing the actual command output that reveals source files, examples, test scripts, documentation, and other repository components.

Through multiple rounds of repository analysis, the agent develops a comprehensive understanding of the project structure and dependencies. Once sufficient context is gathered, the agent begins the translation process by identifying critical components that require translation, such as core functionality modules. The agent employs a thoughtful approach where it reasons about implementation requirements and develops targeted translation strategies. For example, when encountering a function like \texttt{busy\_wait\_milliseconds}, the agent uses \texttt{CreateFile(filepath="...")} to create the corresponding Python file and implements the function while preserving the original functionality and adapting it appropriately for the target language ecosystem.

Throughout the entire translation process, the agent maintains contextual awareness of the project structure, ensuring that translated components integrate seamlessly with the overall repository architecture. The agent leverages its action space in an iterative manner, continuously reading source files to understand implementation details, creating translated files with language-appropriate adaptations, executing commands to validate translation correctness, and refining the implementation based on execution feedback until the translation task is completed or a timeout occurs.

\section{Experimental Setup}

\subsection{Research Questions}

We aim to answer the following key research questions (RQs) that explore the utility of \toolname~ and RepoTransAgent:

\begin{itemize}[left=10pt]
    \item \textbf{RQ1 (Performance of LLM-based Translation Methods):} How do different LLM-based translation methods perform in repository-level code translation tasks?
    \item \textbf{RQ2 (Performance Differences Across translation pairs):} What are the performance differences across different programming translation pairs?
    \item \textbf{RQ3 (Impact of Dependency Complexity):} How does dependency complexity affect the difficulty and performance of repository-level code translation?
    \item \textbf{RQ4 (Error Analysis):} What are the main types of errors that occur in repository-level code translation and what are their underlying causes?
\end{itemize}

\subsection{Model Selection}
\label{sec:ModelSelection}
As shown in Table~\ref{table:models}, we select 8 advanced LLMs from four different companies as our subject LLMs, which include 4 open-source and 4 closed-source LLMs representing the state-of-the-art in large language model capabilities.

For open-source LLMs, we include Alibaba's Qwen3-235B-A22B and its reasoning variant Qwen3-235B-A22B-think, both released in April 2025 with 235B parameters and 32K context windows. These models represent Alibaba's latest advancement in large-scale language modeling. Additionally, we evaluate DeepSeek's two flagship models: DeepSeek-Chat and DeepSeek-Reasoner, both featuring 236B parameters with extended 128K context windows. DeepSeek-Chat was released in March 2025, while DeepSeek-Reasoner, released in May 2025, incorporates enhanced reasoning capabilities that have demonstrated superior performance on complex reasoning tasks.

For closed-source LLMs, we examine four leading models from major AI companies. From Anthropic, we evaluate Claude-Sonnet-4, released in May 2025 with a 200K context window, representing one of the most capable reasoning models available. Google's Gemini-2.5-Flash-Lite, released in June 2025 with a 64K context window, offers efficient performance with reduced computational requirements. From OpenAI, we include two models: GPT-4.1, released in April 2025 with an impressive 1M context window enabling processing of extremely long documents, and o3-mini, released in January 2025 with a 200K context window, designed for efficient reasoning tasks.

Notably, most of the LLMs used in our experiments support substantial context windows ranging from 32K to 1M tokens, which significantly aids in understanding and processing long sequences of text or code. This extended context capability is particularly valuable for tasks requiring comprehensive understanding of large-scale content or complex multi-step reasoning.

\begin{table*}[t]
  \centering
  \footnotesize
  \setlength\tabcolsep{8pt}
  \caption{The selection of backbone LLMs.}
  \renewcommand{\arraystretch}{1.3}
\resizebox{0.85\linewidth}{!}{
    \begin{tabular}{|c|c|c|c|c|c|}
    \hline
    \rowcolor[gray]{0.95}
    \textbf{Source} & \textbf{Model Name} & \textbf{Company} & \textbf{Size} & \textbf{Context Window} & \textbf{Release Date} \\
    \hline
    \hline
    \multirow{4}{*}{\makecell{\textbf{Open}\\\textbf{Source}}} 
    & Qwen3-235B-A22B & Alibaba & 235B & 32K & Apr 2025 \\
    & Qwen3-235B-A22B-think & Alibaba & 235B & 32K & Apr 2025 \\
    & DeepSeek-Chat & DeepSeek & 236B & 128K & Mar 2025 \\
    & DeepSeek-Reasoner & DeepSeek & 236B & 128K & May 2025 \\
    \hline
    \multirow{4}{*}{\makecell{\textbf{Closed}\\\textbf{Source}}} 
    & Claude-Sonnet-4 & Anthropic & - & 200K & May 2025 \\
    & Gemini-2.5-Flash-Lite & Google & - & 64K & Jun 2025 \\
    & GPT-4.1 & OpenAI & - & 1M & Apr 2025 \\
    & o3-mini & OpenAI & - & 200K & Jan 2025 \\
    \hline
    \end{tabular}}%
 \label{table:models}
\end{table*}

\subsection{Evaluation Metrics}
\label{EvaluationMetrics}

We evaluate the translation and debugging performance by the following metrics:

\begin{itemize}[left=10pt]
    \item \textbf{SR} (Success Rate): The metric SR measures the percentage of translation tasks that successfully \textbf{pass all test cases}. A translation task is considered successful if and only if all test cases in the task are passed.
    
    Let $T_i$ represent the number of test cases that pass for the $i$-th translation task, and $N_i$ represent the total number of test cases for the $i$-th translation task. The success indicator for the $i$-th task is defined as:
    
    \[
    \small
    S_i = \begin{cases}
    1 & \text{if } T_i = N_i \\
    0 & \text{otherwise}
    \end{cases}
    \tag{1}
    \]
    
    The Success Rate is calculated as:
    
    \[
    \small
    \text{SR} = \frac{1}{R} \sum_{i=1}^{R} S_i
    \tag{2}
    \]
    
    where $R$ is the total number of translation tasks (repositories) in the benchmark.

    \item \textbf{CR} (Compilation Rate): The metric CR measures the percentage of translation tasks that \textbf{successfully compile} without any compilation errors.
    
    Let $C_i$ represent the compilation indicator for the $i$-th translation task:
    
    \[
    \small
    C_i = \begin{cases}
    1 & \text{if the } i\text{-th task compiles successfully} \\
    0 & \text{otherwise}
    \end{cases}
    \tag{3}
    \]
    
    The Compilation Rate is calculated as:
    
    \[
    \small
    \text{CR} = \frac{1}{R} \sum_{i=1}^{R} C_i
    \tag{4}
    \]
    
    where $R$ is the total number of translation tasks in the benchmark.

    \item \textbf{APR} (Average Pass Rate): The metric APR measures the \textbf{average percentage of test cases passed} across all translation tasks. It reflects the fine-grained performance at the individual test case level.

    \[
    \small
    \text{APR} = \frac{1}{R} \sum_{i=1}^{R} \frac{T_i}{N_i}
    \tag{5}
    \]
    
    where $T_i$ represents the number of passed test cases for the $i$-th translation task, $N_i$ represents the total number of test cases for the $i$-th translation task, and $R$ is the total number of translation tasks.

    \item \textbf{AMPR} (Average Module Pass Rate): The metric AMPR measures the \textbf{average percentage of test modules passed} across all translation tasks. A test module is considered passed if and only if all test cases within that module are passed.
    
    Let $M_i$ represent the total number of test modules for the $i$-th translation task, and $P_i$ represent the number of modules that pass all their test cases. The module pass indicator for the $j$-th module in the $i$-th task is defined as:
    
    \[
    \small
    M_{i,j} = \begin{cases}
    1 & \text{if all test cases in module } j \text{ of task } i \text{ pass} \\
    0 & \text{otherwise}
    \end{cases}
    \tag{6}
    \]
    
    Then $P_i = \sum_{j=1}^{M_i} M_{i,j}$, and the Average Module Pass Rate is calculated as:
    
    \[
    \small
    \text{AMPR} = \frac{1}{R} \sum_{i=1}^{R} \frac{P_i}{M_i}
    \tag{7}
    \]
    
    where $R$ is the total number of translation tasks in the benchmark.
    
\end{itemize}

It is worth noting that many previous works use \textbf{similarity-based metrics} like \textit{BLEU}~\cite{papineni2002bleu}, \textit{CodeBLEU}~\cite{ren2020codebleu}, or other metrics which calculate the overlapping tokens between references and translations~\cite{nguyen2013lexical,karaivanov2014phrase,barone2017parallel,aggarwal2015using,yan2023codetransocean} to evaluate the quality of translations. However, similarity-based metrics ignore the \textbf{syntactic correctness} and \textbf{functional correctness} of translations~\cite{lachaux2020transcoder}. On the one hand, translations with high similarity to the reference cannot avoid having grammatical errors and functional errors. On the other hand, equivalent programs with different implementations may have low similarity. Therefore, we decide to use \textbf{execution-based metrics} mentioned above to evaluate the performance.

\vspace{-10pt}
\subsection{Execution Environment}  
To prevent LLMs from generating malicious code that could execute on the local machine and cause damage, we run the generated code in a sandbox (an isolated environment). 
We use \textit{Docker}~\cite{docker} as our code execution space, and bridge Docker's network with the local machine, allowing it to access the internet to ensure the dependencies in the ``pom.xml'' file can be successfully installed.

\section{Evaluation Results}
\label{sec:EvaluationResults}
\subsection{RQ1: Performance of LLM-based Translation Methods}
\label{sec:PerformanceOfTranslation}

\begin{table*}[t]
  \centering
  \footnotesize
  \setlength\tabcolsep{4pt}
  \caption{Performance evaluation results across different LLMs and methods.}
  \renewcommand{\arraystretch}{1.3}
\resizebox{0.9\linewidth}{!}{
    \begin{tabular}{|l|c|c|c|c||l|c|c|c|c|}
    \hline
    \rowcolor[gray]{0.95}
    \textbf{Method} & \textbf{SR} & \textbf{CR} & \textbf{APR} & \textbf{AMPR} & \textbf{Method} & \textbf{SR} & \textbf{CR} & \textbf{APR} & \textbf{AMPR} \\
    \hline
    \hline
    TranslationOnly$_{\text{Qwen3}}$ & 0.0\% & 26.2\% & 18.6\% & 16.2\% & TranslationOnly$_{\text{Claude}}$ & 0.0\% & 28.0\% & 16.4\% & 14.7\% \\
    ErrorFeedback$_{\text{Qwen3}}$ & 12.4\% & 30.5\% & 23.0\% & 20.5\% & ErrorFeedback$_{\text{Claude}}$ & 11.3\% & 37.5\% & 26.8\% & 24.3\% \\
    RepoTransAgent$_{\text{Qwen3}}$ & 16.9\% & 34.4\% & 26.4\% & 23.6\% & RepoTransAgent$_{\text{Claude}}$ & 32.8\% & 54.8\% & 44.8\% & 41.3\% \\
    \hline
    TranslationOnly$_{\text{Qwen3-think}}$ & 0.0\% & 25.9\% & 18.7\% & 16.7\% & TranslationOnly$_{\text{Gemini}}$ & 0.0\% & 32.5\% & 6.1\% & 5.2\% \\
    ErrorFeedback$_{\text{Qwen3-think}}$ & 13.8\% & 30.9\% & 22.9\% & 20.8\% & ErrorFeedback$_{\text{Gemini}}$ & 4.1\% & 31.2\% & 10.3\% & 9.2\% \\
    RepoTransAgent$_{\text{Qwen3-think}}$ & 19.1\% & 36.0\% & 27.3\% & 25.0\% & RepoTransAgent$_{\text{Gemini}}$ & 11.3\% & 34.4\% & 21.6\% & 19.9\% \\
    \hline
    TranslationOnly$_{\text{DeepSeek}}$ & 0.0\% & 27.0\% & 17.2\% & 15.2\% & TranslationOnly$_{\text{GPT-4.1}}$ & 0.0\% & 26.5\% & 19.6\% & 17.4\% \\
    ErrorFeedback$_{\text{DeepSeek}}$ & 13.9\% & 30.2\% & 24.3\% & 21.6\% & ErrorFeedback$_{\text{GPT-4.1}}$ & 15.6\% & 37.7\% & 29.0\% & 26.1\% \\
    RepoTransAgent$_{\text{DeepSeek}}$ & 22.5\% & 36.5\% & 30.4\% & 27.9\% & RepoTransAgent$_{\text{GPT-4.1}}$ & 32.8\% & 53.3\% & 45.4\% & 40.8\% \\
    \hline
    TranslationOnly$_{\text{DeepSeek-R}}$ & 0.0\% & 10.9\% & 1.5\% & 1.3\% & TranslationOnly$_{\text{o3-mini}}$ & 0.2\% & 33.3\% & 6.0\% & 5.4\% \\
    ErrorFeedback$_{\text{DeepSeek-R}}$ & 0.9\% & 10.8\% & 2.3\% & 1.9\% & ErrorFeedback$_{\text{o3-mini}}$ & 8.7\% & 38.5\% & 11.4\% & 10.8\% \\
    RepoTransAgent$_{\text{DeepSeek-R}}$ & 1.2\% & 10.9\% & 2.5\% & 2.1\% & RepoTransAgent$_{\text{o3-mini}}$ & 12.0\% & 41.0\% & 24.2\% & 23.0\% \\
    \hline
    \end{tabular}}%
 \label{table:performanceResults}
\end{table*}

Table~\ref{table:performanceResults} presents comprehensive evaluation results across different LLMs and methodologies on our RepoTransBench benchmark. The results reveal significant variations in repository-level code translation capabilities and highlight the effectiveness of our proposed approach.

The results demonstrate that repository-level code translation remains a challenging task for current LLMs. Even the best-performing configuration (RepoTransAgent with Claude and GPT-4.1) achieves only 32.8\% success rate, indicating that existing models struggle significantly with the complexity of entire software repositories. However, our RepoTransAgent framework consistently demonstrates superior performance across all evaluated models. For instance, with Claude, RepoTransAgent achieves 32.8\% SR compared to 0.0\% for TranslationOnly and 11.3\% for ErrorFeedback, while with GPT-4.1, it achieves 32.8\% SR versus 0.0\% and 15.6\% respectively. This consistent improvement across different LLMs validates the effectiveness of our agent-based approach for repository-level code translation.

\begin{myboxc} \textbf{Finding 1: }
Our RepoTransAgent framework consistently outperforms baseline approaches across all evaluated backbone models.
However, repository-level code translation remains highly challenging for current LLM-based methods, with the best-performing method achieving only a 32.8\% success rate. 
\end{myboxc}

The evaluation reveals a substantial performance gap between leading closed-source models and open-source alternatives. Claude and GPT-4.1 demonstrate superior performance with RepoTransAgent achieving 32.8\% SR for both models, significantly outperforming open-source models such as Qwen3 (16.9\%), DeepSeek-Chat (22.5\%), and Gemini (11.3\%). Notably, reasoning-focused models do not show clear advantages over their standard counterparts. Qwen3-think achieves 19.1\% compared to Qwen3's 16.9\%, representing only a modest improvement. More surprisingly, DeepSeek-Reasoner performs dramatically worse than DeepSeek-Chat (1.2\% vs 22.5\%), which can be attributed to its excessively long reasoning chains that lead to context management difficulties, request timeouts, and other technical issues that hinder effective translation performance.

\begin{myboxc} \textbf{Finding 2: }
Open-source models generally underperform leading closed-source models (Claude and GPT-4.1). Reasoning-based models exhibit no clear advantages, with DeepSeek-Reasoner performing significantly worse than DeepSeek-Chat due to context management and request timeout issues resulting from overly long reasoning chains.
\end{myboxc}

An important observation is that direct translation approaches (TranslationOnly) consistently fail to produce repositories that pass all test cases, with 0.0\% success rate across most models (except o3-mini with 0.2\%). However, these approaches can still achieve partial success, with compilation rates ranging from 10.9\% (DeepSeek-Reasoner) to 33.3\% (o3-mini), and some test cases passing as evidenced by non-zero APR values. For example, TranslationOnly with Claude achieves 28.0\% compilation rate and 16.4\% average pass rate, indicating that while complete functional correctness is extremely difficult to achieve through direct translation, partial syntactic correctness and limited functional correctness are attainable.

\begin{myboxc} \textbf{Finding 3: }
The translation-only method rarely produces repositories that pass all test cases. However, it can still achieve partial success with some repositories being compilable and passing individual test cases, highlighting the gap between syntactic and complete functional correctness.
\end{myboxc}


\subsection{RQ2: Performance Differences Across translation pairs}  
\label{PerformanceOfLanguagePairs}

\begin{table*}[t]
  \centering
  \footnotesize
  \setlength\tabcolsep{4pt}
  \caption{Performance of RepoTransAgent across different translation pairs.}
  \renewcommand{\arraystretch}{1.2}
\resizebox{\linewidth}{!}{
    \begin{tabular}{|l|c|c|c|c|c|c|c|c|c|c|c|c|c|c|c|c|}
    \hline
    \rowcolor[gray]{0.95}
    \multirow{2}{*}[0.5\tabcolsep]{\textbf{Translation Pair}} & \multicolumn{4}{c|}{\textbf{RepoTransAgent$_{\text{Qwen3}}$}} & \multicolumn{4}{c|}{\textbf{RepoTransAgent$_{\text{Qwen3-think}}$}} & \multicolumn{4}{c|}{\textbf{RepoTransAgent$_{\text{DeepSeek}}$}} & \multicolumn{4}{c|}{\textbf{RepoTransAgent$_{\text{DeepSeek-R}}$}} \\
    \cline{2-17}
    \rowcolor[gray]{0.95}
    & \textbf{SR} & \textbf{CR} & \textbf{APR} & \textbf{AMPR} & \textbf{SR} & \textbf{CR} & \textbf{APR} & \textbf{AMPR} & \textbf{SR} & \textbf{CR} & \textbf{APR} & \textbf{AMPR} & \textbf{SR} & \textbf{CR} & \textbf{APR} & \textbf{AMPR} \\
    \hline
    \hline
    C\#$\rightarrow$Java & 8.2 & 8.2 & 8.2 & 8.2 & 9.3 & 9.3 & 9.3 & 9.3 & 9.3 & 10.3 & 10.3 & 10.3 & 3.1 & 3.1 & 3.1 & 3.1 \\
    \hline
    C++$\rightarrow$Python & 58.0 & 85.6 & 76.0 & 71.0 & 50.3 & 83.4 & 71.3 & 66.2 & 55.8 & 81.8 & 72.7 & 69.1 & 0.6 & 5.0 & 1.5 & 1.5 \\
    \hline
    C$\rightarrow$Python & 43.4 & 91.0 & 63.1 & 56.3 & 33.6 & 84.4 & 62.0 & 52.6 & 43.4 & 80.3 & 62.3 & 56.1 & 1.6 & 5.7 & 1.6 & 1.6 \\
    \hline
    C$\rightarrow$Rust & 6.6 & 10.7 & 9.7 & 8.2 & 12.3 & 14.8 & 13.7 & 13.1 & 23.8 & 23.8 & 23.8 & 23.8 & 0.0 & 0.0 & 0.0 & 0.0 \\
    \hline
    JavaScript$\rightarrow$Python & 28.0 & 82.0 & 48.2 & 40.5 & 24.3 & 77.8 & 50.5 & 39.1 & 32.8 & 79.4 & 55.4 & 45.5 & 8.5 & 90.5 & 19.4 & 15.7 \\
    \hline
    Java$\rightarrow$C\# & 0.0 & 8.9 & 0.0 & 0.0 & 0.7 & 4.1 & 0.7 & 0.7 & 5.5 & 7.5 & 7.5 & 7.5 & 0.0 & 0.0 & 0.0 & 0.0 \\
    \hline
    Java$\rightarrow$Go & 17.1 & 20.5 & 19.5 & 17.8 & 12.3 & 13.7 & 13.5 & 13.0 & 19.9 & 24.0 & 23.4 & 19.9 & 0.0 & 0.0 & 0.0 & 0.0 \\
    \hline
    Java$\rightarrow$Python & 43.8 & 80.8 & 70.3 & 67.8 & 43.2 & 82.9 & 70.2 & 67.6 & 50.7 & 80.8 & 71.9 & 69.1 & 0.7 & 6.2 & 2.0 & 2.0 \\
    \hline
    Matlab$\rightarrow$Python & 21.9 & 59.4 & 39.4 & 34.2 & 15.6 & 59.4 & 36.1 & 31.1 & 26.6 & 53.1 & 39.0 & 36.9 & 0.0 & 12.5 & 0.0 & 0.0 \\
    \hline
    Python$\rightarrow$C++ & 0.6 & 3.5 & 0.6 & 0.6 & 0.6 & 7.0 & 0.6 & 0.6 & 1.2 & 4.7 & 1.2 & 1.2 & 0.0 & 0.0 & 0.0 & 0.0 \\
    \hline
    Python$\rightarrow$Go & 12.3 & 14.0 & 13.4 & 12.9 & 9.4 & 9.9 & 9.9 & 9.4 & 14.0 & 18.1 & 17.5 & 15.2 & 0.0 & 0.0 & 0.0 & 0.0 \\
    \hline
    Python$\rightarrow$Java & 0.6 & 1.2 & 1.2 & 1.2 & 1.2 & 1.2 & 1.7 & 1.2 & 1.8 & 2.9 & 2.3 & 1.8 & 0.0 & 0.0 & 0.0 & 0.0 \\
    \hline
    Python$\rightarrow$Rust & 5.3 & 5.8 & 5.8 & 5.8 & 4.7 & 4.7 & 4.7 & 4.7 & 8.8 & 8.8 & 8.8 & 8.8 & 0.0 & 0.0 & 0.0 & 0.0 \\
    \hline
    \hline
    \rowcolor[gray]{0.95}
    \multirow{2}{*}[0.5\tabcolsep]{\textbf{Translation Pair}} & \multicolumn{4}{c|}{\textbf{RepoTransAgent$_{\text{Claude}}$}} & \multicolumn{4}{c|}{\textbf{RepoTransAgent$_{\text{Gemini}}$}} & \multicolumn{4}{c|}{\textbf{RepoTransAgent$_{\text{GPT-4.1}}$}} & \multicolumn{4}{c|}{\textbf{RepoTransAgent$_{\text{o3-mini}}$}} \\
    \cline{2-17}
    \rowcolor[gray]{0.95}
    & \textbf{SR} & \textbf{CR} & \textbf{APR} & \textbf{AMPR} & \textbf{SR} & \textbf{CR} & \textbf{APR} & \textbf{AMPR} & \textbf{SR} & \textbf{CR} & \textbf{APR} & \textbf{AMPR} & \textbf{SR} & \textbf{CR} & \textbf{APR} & \textbf{AMPR} \\
    \hline
    \hline
    C\#$\rightarrow$Java & 28.9 & 34.0 & 33.0 & 33.0 & 8.2 & 10.3 & 11.1 & 10.3 & 20.6 & 22.7 & 25.3 & 22.7 & 3.1 & 4.1 & 4.1 & 4.1 \\
    \hline
    C++$\rightarrow$Python & 63.0 & 97.8 & 83.2 & 80.4 & 21.0 & 84.5 & 56.4 & 51.9 & 55.2 & 96.1 & 78.3 & 75.4 & 3.9 & 99.4 & 55.1 & 51.5 \\
    \hline
    C$\rightarrow$Python & 61.5 & 98.4 & 79.7 & 72.8 & 37.7 & 77.0 & 56.3 & 50.0 & 54.9 & 90.2 & 76.9 & 68.5 & 37.7 & 87.7 & 57.6 & 52.4 \\
    \hline
    C$\rightarrow$Rust & 47.5 & 57.4 & 54.3 & 48.8 & 22.1 & 27.0 & 25.2 & 24.6 & 47.5 & 59.0 & 54.0 & 49.2 & 67.2 & 70.5 & 69.1 & 68.2 \\
    \hline
    JavaScript$\rightarrow$Python & 53.4 & 96.3 & 74.8 & 67.4 & 3.2 & 95.8 & 29.3 & 24.4 & 43.4 & 86.2 & 68.5 & 60.0 & 34.9 & 94.7 & 47.7 & 43.9 \\
    \hline
    Java$\rightarrow$C\# & 19.2 & 30.8 & 24.0 & 24.0 & 0.0 & 1.4 & 0.7 & 0.7 & 21.2 & 23.3 & 25.5 & 25.2 & 0.7 & 1.4 & 0.7 & 0.7 \\
    \hline
    Java$\rightarrow$Go & 36.3 & 43.8 & 43.0 & 36.3 & 14.4 & 16.4 & 16.4 & 16.4 & 33.6 & 47.3 & 45.7 & 34.2 & 4.1 & 4.1 & 4.1 & 4.1 \\
    \hline
    Java$\rightarrow$Python & 45.9 & 98.6 & 79.8 & 77.0 & 38.4 & 86.3 & 64.1 & 62.7 & 58.2 & 86.3 & 79.8 & 77.8 & 3.4 & 97.3 & 56.3 & 54.7 \\
    \hline
    Matlab$\rightarrow$Python & 48.4 & 95.3 & 67.8 & 65.4 & 14.1 & 35.9 & 26.0 & 22.8 & 34.4 & 71.9 & 59.8 & 57.7 & 4.7 & 96.9 & 20.3 & 19.7 \\
    \hline
    Python$\rightarrow$C++ & 3.5 & 26.3 & 6.2 & 4.1 & 0.0 & 1.2 & 0.7 & 0.0 & 9.9 & 38.0 & 11.1 & 10.5 & 0.6 & 0.6 & 0.6 & 0.6 \\
    \hline
    Python$\rightarrow$Go & 18.7 & 31.6 & 29.3 & 23.4 & 1.2 & 1.2 & 1.2 & 1.2 & 19.9 & 33.3 & 31.4 & 21.1 & 1.2 & 1.2 & 1.2 & 1.2 \\
    \hline
    Python$\rightarrow$Java & 5.8 & 8.8 & 8.2 & 8.2 & 0.0 & 0.0 & 0.0 & 0.0 & 7.0 & 7.0 & 9.0 & 7.0 & 1.8 & 1.8 & 1.8 & 1.8 \\
    \hline
    Python$\rightarrow$Rust & 11.7 & 17.5 & 17.1 & 15.2 & 1.2 & 1.8 & 1.8 & 1.8 & 26.3 & 35.7 & 34.5 & 32.7 & 1.8 & 1.8 & 1.8 & 1.8 \\
    \hline
    \end{tabular}}%
 \label{table:translationPairs}
\end{table*}

Table~\ref{table:translationPairs} presents detailed performance analysis of RepoTransAgent across 13 different translation pairs and multiple LLMs. The results reveal significant insights about the inherent difficulties and characteristics of different programming language translation combinations.

The most striking pattern in our results is the fundamental asymmetry between translating from statically-typed languages to dynamically-typed languages versus the reverse direction. Translations from static languages (C, C++, Java) to Python consistently achieve high success rates across most models: Claude achieves 61.5\% for C$\rightarrow$Python, 63.0\% for C++$\rightarrow$Python, and 45.9\% for Java$\rightarrow$Python, while GPT-4.1 reaches 54.9\%, 55.2\%, and 58.2\% respectively. This success stems from the fact that static languages provide explicit type information, memory management details, and structured interfaces that can be effectively simplified and adapted to Python's more flexible paradigm.

Conversely, Python-to-static-language translations face severe challenges across all evaluated models. Python$\rightarrow$Java achieves only 0.0-7.0\% success rates, Python$\rightarrow$C++ reaches merely 0.0-9.9\%, and Python$\rightarrow$Rust performs between 1.2-26.3\%. This dramatic performance gap reflects the fundamental challenge of inferring static type information, memory management strategies, and explicit interface definitions from Python's dynamic and implicit programming model. The translation process must essentially reverse-engineer the implicit contracts and assumptions present in dynamically-typed code.

\begin{myboxc} \textbf{Finding 4: }
Translation from statically-typed languages to dynamically-typed achieves substantially higher success rates (45-63\%) compared to the reverse direction (typically below 10\%), revealing a fundamental asymmetry in translation difficulty due to the challenges of inferring explicit type and interface information from dynamic code.
\end{myboxc}

Our analysis reveals that different LLM backbones exhibit surprising specialization patterns for specific translation pairs, likely reflecting their training data composition and architectural biases. Most notably, o3-mini demonstrates exceptional performance on C$\rightarrow$Rust translation with 67.2\% success rate, significantly outperforming leading models like Claude (47.5\%) and GPT-4.1 (47.5\%) on the same pair. This suggests that o3-mini's training included substantial Rust-related content or system-level programming examples that enhanced its understanding of memory safety patterns and ownership concepts critical for C-to-Rust translation.

Similarly, we observe that certain models excel at specific paradigm shifts while struggling with others. For instance, Claude shows strong performance across most translation pairs but particularly excels at translating to Python (achieving the highest success rates for C$\rightarrow$Python, C++$\rightarrow$Python, and Matlab$\rightarrow$Python), suggesting robust training on Python codebases. In contrast, Gemini demonstrates highly inconsistent performance, achieving strong compilation rates for some pairs (95.8\% for JavaScript$\rightarrow$Python compilation) but near-zero functional success rates, indicating potential gaps in understanding semantic equivalence across languages. These patterns suggest that training data distribution and architectural choices significantly influence model performance on specific language combinations.

\begin{myboxc} \textbf{Finding 5: }
Different LLM backbones demonstrate specialized advantages for specific translation pairs (e.g., o3-mini excelling at C$\rightarrow$Rust with 67.2\% vs. Claude's 47.5\%), likely reflecting training data composition and architectural biases that favor particular programming languages.
\end{myboxc}


\subsection{RQ3: Impact of Dependency Complexity}

\begin{figure*}
    \centering
    \includegraphics[width=\linewidth]{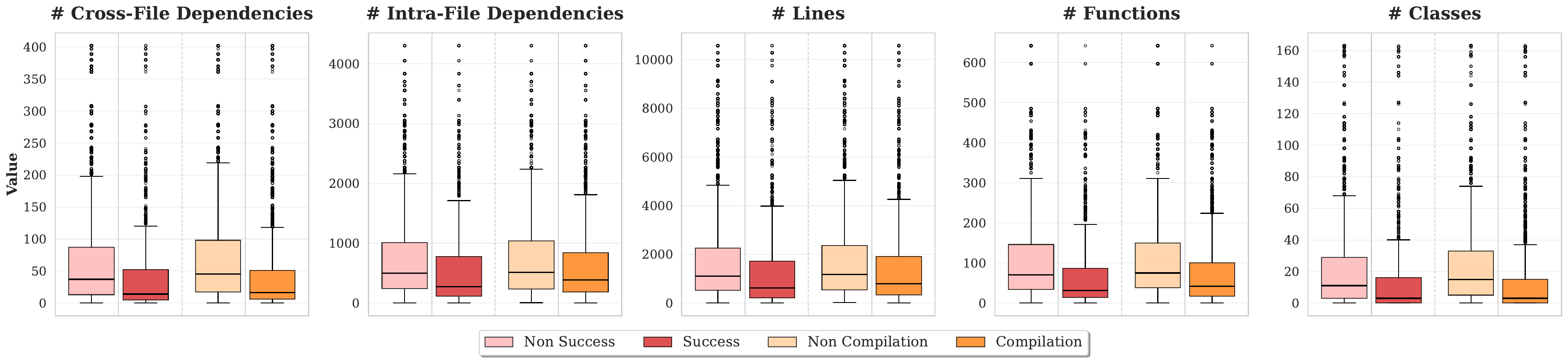}
    \caption{The effect of code length and functional complexity on translation performance. The light colors (\textcolor[HTML]{ffb3b3}{\rule{0.5em}{0.5em}} \textcolor[HTML]{ffcc99}{\rule{0.5em}{0.5em}}) represent non-successful outcomes, while dark colors (\textcolor[HTML]{d62728}{\rule{0.5em}{0.5em}} \textcolor[HTML]{ff7f0e}{\rule{0.5em}{0.5em}}) represent successful outcomes. The first two groups (\textcolor[HTML]{ffb3b3}{\rule{0.5em}{0.5em}} \textcolor[HTML]{d62728}{\rule{0.5em}{0.5em}}) in each subplot show success/non-success results, while the last two groups (\textcolor[HTML]{ffcc99}{\rule{0.5em}{0.5em}} \textcolor[HTML]{ff7f0e}{\rule{0.5em}{0.5em}}) show compilation/non-compilation results. To improve clarity, outliers where metrics exceed the 95th percentile have been omitted.}
    \label{fig:AffectOfComplexity}
\end{figure*}

Figure~\ref{fig:AffectOfComplexity} presents a comprehensive analysis of how repository complexity characteristics affect translation performance across different dimensions. We examine the relationship between various complexity metrics and translation outcomes to understand the fundamental challenges posed by repository-level code translation.

The analysis reveals a clear inverse relationship between repository complexity and translation success across multiple dimensions. For cross-file dependencies, successful translations consistently exhibit lower median values compared to failed translations, indicating that repositories with fewer inter-module dependencies are more amenable to successful translation. Similarly, intra-file dependencies show that successful translations tend to have simpler internal dependency structures. The pattern extends to basic size metrics, where successful translations typically involve repositories with fewer lines of code, fewer functions, and fewer classes. This consistent trend across all complexity dimensions suggests that current LLMs struggle systematically with increased repository complexity, regardless of whether the complexity stems from structural dependencies, code volume, or functional richness.

\begin{myboxc} \textbf{Finding 6: }
Repository complexity across all dimensions (cross-file dependencies, intra-file dependencies, code length, and structural complexity) inversely correlates with translation success, indicating that current LLMs struggle systematically with complex repository structures.
\end{myboxc}

\subsection{RQ4: Error Analysis}

We conduct an error analysis across three rounds of experiments, ultimately identifying common errors in repository-level code translation. These errors are classified into five categories: E1 (Configuration File Issues), E2 (Limited Understanding Ability Issues), E3 (Incomplete Generation Issues), E4 (Language Feature Issues), E5 (Encoding Issues). Due to space constraints, we provide one case for each category.

\begin{figure}
    \centering
    \includegraphics[width=\linewidth]{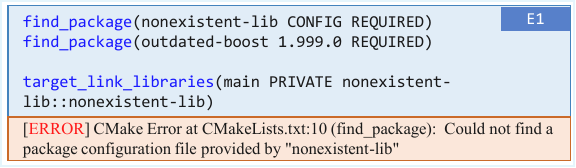}
    \caption{Error Type 1: Configuration File Issues.}
    \label{fig:ErrorType1}
\end{figure}

\textbf{E1. Configuration File Issues} often arise when build-related content (e.g., ``CMakeLists.txt'' in a C++ CMake project) is not configured correctly. Figure~\ref{fig:ErrorType1} shows an error due to an unresolved dependency as the package ``nonexistent-lib'' cannot be found through CMake's package configuration system.
Beyond this common error, we observe other configuration-related issues including version compatibility problems, missing dependency declarations in build files (e.g., ``package.json'', ``requirements.txt'', ``Cargo.toml''), and platform-specific configuration errors. Some LLMs occasionally generate inappropriate content for configuration files, leading to build failures.

\secmargin
\begin{myboxc} \textbf{Finding 7: } 
Unlike fine-grained code translation works, repository-level code translation requires proper configuration of files such as the “CMakeLists.txt” file in C++ repositories. This can lead to dependency-related issues (e.g., non-existent dependencies, version mismatches, etc.). Solving these problems requires a clear understanding of the related calls and the latest dependencies.
\end{myboxc} 
\secmargin

\begin{figure}
    \centering
    \includegraphics[width=\linewidth]{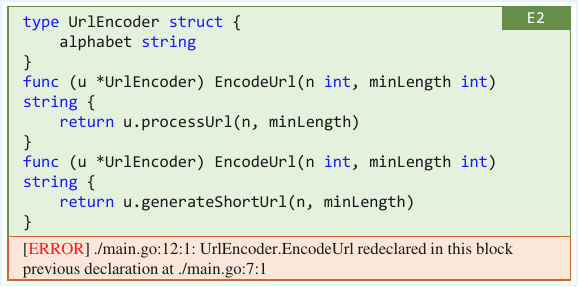}
    \figmargin
    \caption{Error Type 2: Limited Understanding Ability Issues.}
    \figmargin
    \label{fig:ErrorType2}
\end{figure}

\textbf{E2. Limited Understanding Ability Issues} usually arise due to unfamiliarity with the code context during translation.
Figure~\ref{fig:ErrorType2} shows two methods with identical names and parameter signatures in a Go struct, leading to a redeclaration error. This occurs because the previously generated method is overlooked when generating new functions, often due to insufficient context awareness in long codebases.
Beyond this example, we observe other context-related issues including incorrect function calls with mismatched argument types, improper imports due to misunderstanding repository structure, and variable scope conflicts. These errors typically stem from LLMs' limited ability to maintain comprehensive understanding of the entire codebase during translation.

\secmargin
\begin{myboxc} \textbf{Finding 8: } 
Due to the long code length and complex repository structure, a lack of understanding of the repository context may result in generating inappropriate code. To mitigate these issues, taking measures to enhance the focus on relevant context within the repository might be helpful.
\end{myboxc} 

\secmargin

\begin{figure}
    \centering
    \includegraphics[width=\linewidth]{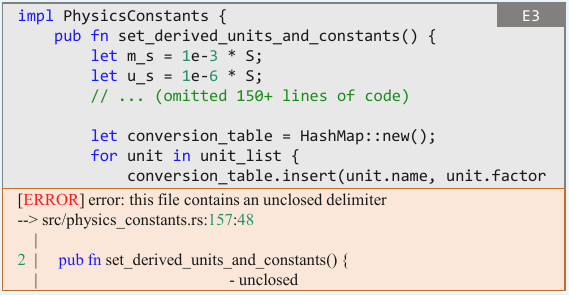}
    \figmargin
    \caption{Error Type 3: Incomplete Generation Issues.}
    \figmargin
    \label{fig:ErrorType3}
\end{figure}

\textbf{E3. Incomplete Generation Issues} often occur due to LLMs' limitation in instruction following and code generation abilities. 
Figure~\ref{fig:ErrorType3} shows an example where the LLM fails to complete a long Rust function, leaving unclosed delimiters and incomplete statements. The generation terminates abruptly in the middle of a \texttt{HashMap} insertion, resulting in syntax errors due to missing closing braces and parentheses.
Beyond incomplete function bodies, we observe other generation issues including missing import statements for used packages, incomplete class or struct definitions, and truncated method implementations. These problems often occur when LLMs struggle with long code sequences or when they generate example-style code while omitting essential elements.

\secmargin
\begin{myboxc} \textbf{Finding 9: } 
Some LLMs may struggle to continue generating due to limited instruction following capability, and others may tend to generate example code with some elements omitted. These issues can be automatically identified by dependency analysis and syntax checking. Then we can fix these issues by regenerating or replacing them with stronger LLMs.
\end{myboxc} 
\secmargin

\begin{figure}
    \centering
    \includegraphics[width=\linewidth]{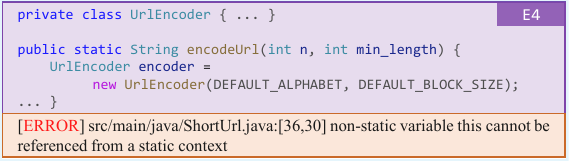}
    \figmargin
    \caption{Error Type 4: Language Feature Issues.}
    \label{fig:ErrorType4}
\end{figure}

\textbf{E4. Language Feature Issues} occur frequently in repository-level code translation due to its functional complexity. 
Figure~\ref{fig:ErrorType4} shows a case where a static method attempts to access a non-static inner class \texttt{UrlEncoder}, which violates Java's static context rules. This error demonstrates the fundamental misunderstanding of static versus non-static member accessibility in Java.
Beyond this example, we observe other language-specific issues including attempts to instantiate abstract classes, direct access to private member variables from external classes, and incorrect use of language-specific keywords or modifiers. These problems typically arise when LLMs perform token-by-token translation without considering the target language's semantic constraints and access control mechanisms.

\secmargin
\begin{myboxc} \textbf{Finding 10: } 
When translating a repository to another language, some LLMs may lack sufficient understanding of language features, leading to related issues. Providing LLMs with more information about the language features in the prompt may help improve the translation performance.
\end{myboxc} 
\secmargin

\begin{figure}
    \centering
    \includegraphics[width=\linewidth]{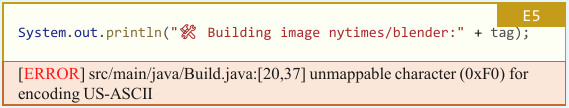}
    \caption{Error Type 5: Encoding Issues.}
    \label{fig:ErrorType5}
\end{figure}

\textbf{E5. Encoding Issues} occur due to incompatible repository encoding formats and can lead to compilation failures when using special characters. Figure~\ref{fig:ErrorType5} shows an example where a Java source file contains an emoji character in a string literal, but the compiler is configured to use US-ASCII encoding, which cannot handle Unicode characters beyond the basic ASCII range.
Beyond emoji usage, we observe other encoding-related issues including problems with non-English characters in comments or string literals, issues when reading files with different encoding formats, and compilation errors in multilingual environments. These problems typically arise when the build system's default encoding configuration is insufficient for the characters present in the translated code.

\secmargin
\begin{myboxc} \textbf{Finding 11: } 
Non-US-ASCII characters commonly occur in repositories, and encoding-related issues can sometimes arise when reading or writing resources. These problems can be resolved by configuring correct encoding format of the repository.
\end{myboxc} 
\secmargin

\section{Threats to Validity}

\noindent \textbf{Internal Threats.}
The first potential internal threat concerns the \textit{scope of evaluated LLMs}. While we evaluate 8 state-of-the-art LLMs across different categories (open-source, closed-source, and reasoning-focused models), the rapidly evolving landscape of LLMs means that newer models may exhibit different performance characteristics. However, our evaluation includes leading models from major providers (OpenAI, Anthropic, Google, Alibaba, DeepSeek) that represent the current state-of-the-art, ensuring broad coverage of existing capabilities. Additionally, we do not apply fine-tuning methods to these LLMs specifically for repository-level translation, which may impact their performance. This limitation is partially mitigated by including specialized code models that have been pre-trained on code-related tasks.
Another potential threat is the \textit{selection of programming translation pairs}. Our benchmark focuses on 13 translation pairs across 7 programming languages, with emphasis on commonly used languages in software development. While this covers major programming paradigms (object-oriented, functional, systems programming), the results may not generalize to less common languages or domain-specific languages. However, our language selection is based on TIOBE rankings and developer survey data, ensuring relevance to real-world translation needs. The inclusion of diverse language combinations (e.g., C $\leftrightarrow$ Python, Java $\leftrightarrow$ Go, Python $\leftrightarrow$ Rust) provides insights into various translation scenarios encountered in practice.

\noindent \textbf{External Threats.}
The primary external threat involves \textit{LLMs' generation variability}. Large language models exhibit inherent randomness in their outputs, which could affect the reproducibility of our results. To mitigate this threat, we employ consistent experimental settings and analyze performance patterns across multiple samples rather than isolated instances. Additionally, our evaluation focuses on objective metrics (compilation success, test passage) that are less susceptible to generation variance compared to subjective quality assessments.
Another potential threat concerns the \textit{comprehensiveness of functional correctness evaluation}. Our evaluation primarily relies on execution-based metrics using existing test suites from source repositories. While passing all test cases provides strong evidence of functional correctness, it may not capture all edge cases or guarantee complete semantic equivalence. However, this approach represents a significant advancement over similarity-based metrics used in previous work, as it directly validates the operational correctness of translated code. Real-world test suites from production repositories provide more realistic evaluation scenarios compared to artificially constructed benchmarks.
The \textit{repository selection and filtering process} may introduce bias toward certain types of projects. Our filtering criteria (star count, language composition, executability) may favor well-maintained, popular repositories while excluding experimental or domain-specific projects. This bias is intentional to ensure benchmark quality and practical relevance, as successful repository-level translation tools should prioritize handling well-structured, maintainable codebases that represent common development scenarios.
Finally, the \textit{temporal validity} of our benchmark presents a consideration, as programming languages, frameworks, and development practices evolve continuously. However, our focus on fundamental language features and well-established frameworks ensures that our findings remain relevant across reasonable time horizons. The automated benchmark construction framework we develop can facilitate future updates and extensions to maintain benchmark currency.

\vspace{-5pt}
\section{Related Work}
Many benchmarks have been introduced to compare the performance of different translation techniques objectively. CoST~\cite{Zhu2022CoST} and XLCost~\cite{zhu2022xlcost} introduce a snippet and function-level code translation benchmark. CodeXGLUE~\cite{lu2021codexglue} includes a dataset for function-level Java-C\# code translation. TransCoder-test is the evaluation dataset for TransCoder~\cite{lachaux2020transcoder}, which includes the function-level code translation on Python, Java, and C++. Some other benchmarks like HumanEval-X~\cite{humanevalx} source from HumanEval~\cite{humaneval} to construct a function-level code translation benchmark. G-TransEval~\cite{jiao2023GTransEval} provides a more fine-grained taxonomy, including token-level, syntax-level, library-level, and algorithm-level, which is part of a function. CodeNet~\cite{puri2021codenet}, Avatar~\cite{ahmad2021avatar}, xCodeEval~\cite{khan2023xcodeeval} CodeScope~\cite{yan2023codescope} and CodeTransOcean~\cite{yan2023codetransocean} introduce file-level code translation benchmarks which source from code contest platforms like codeforces~\cite{Codeforces}, atcoder~\cite{atcoder}, aizu~\cite{aizu}, Google Code Jam~\cite{GoogleCodeJam}, etc. or task solutions websites like samples from .Net~\cite{dotnetsamples}, d2lai~\cite{d2lai}, rosetta code~\cite{rosettacode}, etc. 
Although these benchmarks can evaluate the capabilities of existing code translation techniques to some extent, they cannot evaluate the performance of current techniques on real-world repository-level code translation tasks. 
Recently, pan et al.~\cite{pan2024lost} manually study two open-source repositories (Apache Commons CLI~\cite{apachecommonscli} and Python Click~\cite{click}) and find that current LLMs struggle to complete the translation tasks of entire repositories. However, they do not provide a sufficient number of repositories and corresponding automatic test suites for evaluation. Besides, the resource and configuration files are ignored in this research.
Recently, Pan et al.~\cite{pan2024lost} manually study two open-source repositories (Apache Commons CLI~\cite{apachecommonscli} and Python Click~\cite{click}) and find that current LLMs struggle to complete the translation tasks of entire repositories. 
~\newrevised{AlphaTrans~\cite{ibrahimzada2025alphatrans} is proposed as a neuro-symbolic compositional technique that decomposes repositories into fragments and translates them in reverse call order with GraalVM-based validation. 
TRACY~\cite{gong2025tracy} is a benchmark proposed to evaluate the execution efficiency of function-level code translation.
However, our main contribution is introducing a real-world multilingual benchmark for verifying the equivalence of repository-level code translation.
}

\vspace{-8pt}
\section{Conclusion}
\vspace{-3pt}
This paper addresses the gap between existing fine-grained code translation benchmarks and real-world software development demands by introducing \toolname, a comprehensive repository-level benchmark with 1,897 samples across 13 translation pairs, and RepoTransAgent, an intelligent agent framework based on the ReAct paradigm for systematic repository translation. Our evaluation reveals that repository-level translation remains challenging, with the best-performing method achieving only a 32.8\% success rate. We also observe strong directional asymmetry in translation difficulty (static-to-dynamic achieving 45-63\% vs. reverse direction below 10\%), model-specific advantages for certain translation pairs reflecting training biases, and inverse correlation between repository complexity and translation success. This paper provides the community with both a challenging benchmark and practical guidance for future research.

\vspace{-8pt}
\section*{Acknowledgements}
\vspace{-3pt}
This work is supported by CCF-Huawei Populus Grove Fund CCF-HuaweiSE202403.

\bibliographystyle{IEEEtran}
\bibliography{ref}

\end{document}